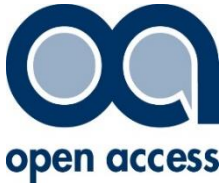
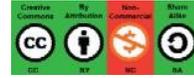

# Location and Type of Crimes in The Philippines: Insights for Crime Prevention and Management

*Liene Leikuma-Rimicane[1]*
Daugavpils University

*Roel F. Ceballos[2]*
University of Southeastern Philippines

*Milton Norman D. Medina[3]\**
Davao Oriental State University

**Abstract**
*The purpose of this study was to determine the association of location and types of crimes in the Philippines and understand the impact of COVID-19 lockdowns by comparing the crime incidence and associations before and during the pandemic. A document review was used as the main method of data collection using the datasets from the Philippine Statistics Authority- Annual Statistical Yearbook (PSA-ASY). The dataset contained the volume of index crimes in the Philippines from 2016 to 2020. The index crimes were broken down into two major categories: crimes against persons and crimes against property. Incidence of crime-by-crime type was available for different administrative regions in the Philippines. Chi-square test and correlation plot of chi-square residual were used to determine the associations between the locations and types of index crimes. A correlation plot of the chi-square residual was used to investigate the patterns of associations. Results suggest that the continuing effort of the Philippine government to fight against criminality has resulted in a steady decline in the incidence of index crimes in the Philippines. The pandemic too contributed to the decline of crime incidence in the country. These results imply that police surveillance activities in highly populated areas and specific interventions to address sexual violence must be in place during community lockdowns. The Philippine National Police should heighten its campaign in violence against women and increase its workforce visibility especially in remote and densely populated areas. The results of this study can be used as input to local government units for developing programs and plans on crime prevention. For future researches, it is recommended to conduct a precinct level analysis for a closer look at crime surveillance.*

---

[1] Faculty of Education and Management, Daugavpils University, Parades Str. 1, Daugavpils, LV-5401, Latvia. Email: liene.rimicane@du.lv, ORCID: 0000-0002-4677-4795
[2] Mathematics and Statistics Department, University of Southeastern Philippines.
Email: roel.ceballos@usep.edu.ph, ORCID: 0000-0001-8267-6482
[3]\* Institute of Agriculture and Life Sciences, Davao Oriental State University, Dahican, Mati City 8200 Philippines; Email: miltonnormanmedina@gmail.com, ORCID: 0000-0001-6858-8048

**22**

**Introduction**

Increasing crime solution efficiency in any nation requires understanding of the crime trends and their associations to specific locations, which will enhance the crime prevention and management of the police and other enforcement agencies. Crime prevention and management are at the forefront of the agenda of the Philippine government under the Duterte administration. One of the administration's goals is to improve the lives of Filipinos by aggressively reducing corruption and crimes (Timberman, 2019). To cite a few of its strategies, the government has implemented the anti-narcotics campaign or most known as the 'War-on-Drugs,' and the continuous fight against criminality, resulting in thousands of drug-peddlers all over the country (Gita-Carlos, 2019).

One step taken by Philippine government towards crime prevention and increase the police visibility was passing of the law to modify the base salary of military and uniformed personnel. This step was also taken to motivate the existing personnel and encourage Filipinos to pursue careers in the Armed Forces of the Philippines. The salary adjustments resulted in a 72.18% increase for all ranks of uniformed personnel (Department of Budget and Management, 2018). As a result, police visibility has increased in different regions as the Philippine National Police (PNP) has also increased the number of police and police stations all over the country. The Philippine National Police have also reported an increase in their overall crime solution efficiency (Gita-Carlos, 2019).

Despite these improvements, the incidence of index crimes remains observable and has not been eliminated in any part of the country. As of February 2022, there were roughly 10 thousand cases of violation of special laws reported in the Philippines. On the other hand, reckless imprudence resulting to damage of property amounted to over seven thousand cases. Physical injury ranks 3rd with 3,753 cases, and reckless imprudence resulting to homicide is the least with 372 cases (Statista, 2022). Patterns and incidence of crimes vary by type and location (Nivette et al., 2021). Hence, there should be a location-specific component in our collective crime prevention and management approach. COVID-19 lockdowns across the globe, mobility restrictions which are collectively called lockdowns, have been implemented to combat the spread of COVID-19 as another component to investigate. Since crime is a social phenomenon, lockdowns have caused shifts in trends and patterns across different locations (Buil-Gil, Zeng, & Kemp, 2021).

Despite the government's effort in reducing the crime rate in the Philippines, there is still numerous recordings of index crimes or crimes against people and property. Since the Philippines is an archipelagic country, it is very important to analyze the type of crimes that occur in specific islands or areas in the country to provide data for efficient police enforcement. Understanding crime trends and their associations to specific locations will enhance the crime prevention and management of the police force. This will lead to an increase in crime solution efficiency. It will also serve as a guide to regional government executives in identifying which crimes to prioritize in their short-term and long-term plans.

Increasing crime solution efficiency requires understanding crime trends and their associations to specific locations, which will enhance the crime prevention and management of the police force. Hence, this study was designed with the objective to examine the patterns of index crimes per location as input to the development plans

**23**

of the different Local Government Units (LGUs) in the Philippines. To achieve this objective, this study aimed to analyze the association of location and types of index crimes in the Philippines and understand the impact of COVID-19 lockdowns by comparing the crime incidence and associations before and during the pandemic. Specifically, this study aimed to determine the predominant index crimes before and during the pandemic across different regions in the Philippines.

**Literature Review**

Crime rates vary greatly from country to country; for example, in 2022, Venezuela ranked with the highest crime rate with 83.58% followed by Papua New Guinea 81.19%, South Africa 77.01% amongst the top three in the world while the Philippines ranked in the 80th place with 42.33% crime index (Balmori de la Miyar, Hoehn-Velasco, & Silverio-Murillo, 2021). Some of the world's lowest crime rates are seen in Switzerland, Denmark, Norway, Japan, and New Zealand (Numbeo, n.d.). There are several reasons for this which include but are not limited to poverty (Chiricos, 1987), unemployment (Fowles & Merva, 1996), and income inequality (Blau & Blau, 1982). The effects of poverty and unemployment are not surprising especially in the Philippines.

The Philippines is one of the 11 Southeast Asian nations with one of the most highly populated countries with approximately 110 million people, second after Indonesia and ranks 13th in the world (www.worldometers.info). The poverty rate of the Philippines has reduced significantly since 2015 from 21.6% to 16.6% in 2021 and a rapid decline of crime rate (14% drop of crime rate) in the country since 2017 under the Duterte Administration. Without a doubt, the campaign against poverty has significantly reduced the crime rate in the country. Moreover, to continue this momentum, the Philippine government has implemented countermeasures in various index crimes in the country.

Interestingly, crime prevention policies have been incorporated in national economic development plans of the Philippines. The Medium-Term Philippine Development Plan embodies as one of its policy frameworks for the improvement of law and order, and law enforcement administration of justice. It emphasizes the government's role to guarantee public safety and national security, while ensuring that the rule of law prevails. Thus, ensuring peace and order rests primarily on the ability of the government to curb criminal activities. In this regard, it is vital to strengthen the criminal justice system. Hence research on the trends of the index crimes and their locations is an important data that will support the criminal justice system in the Philippines especially in the regional and local contexts (Lusthaus et al., 1999).

The occurrence of COVID-19 pandemic shifted the global trend of index crimes (Meyer, Prescott, & Sheng, 2022). It was alarming to find that domestic violence increased during the pandemic (Nivette et al., 2021). In fact, in 2020, various media sources in the United States reported an increase of homicide cases (Asher & Horwitz, 2020; Hilsenrath, 2021; McCarthy, 2020; Struett, 2020). Meanwhile, with the limited opportunities of the criminals due to community lockdowns, other crimes, such as burglary and robbery, were reported to have decreased following the start of the COVID-19 pandemic in the United States (Boman & Gallupe, 2020). Despite there being numerous studies on crime during COVID-19 (Nivette et al., 2021), global research on the association of location and types of index crimes is very limited.

**24**

## Methodology

- *Research Framework*

Crimes in the Philippines are reported based on the location where the crime occurs and the specific crime classification or type. Hence, it is not surprising that crime incidence, as a metric monitored by concerned government agencies, is described based on these characteristics. It is also imperative that crime prevention and management programs of the government will be greatly influenced by the insights produced using this information. Many studies found in literature emphasize that understanding crimes as an input to crime prevention and management requires a thorough investigation between the association of types and locations of crimes (Irvin-Erickson & La Vigne, 2015; Leong & Sung, 2015; Newton & Felson, 2015; Zhou et al., 2021). The framework of the study is provided in Figure 1.

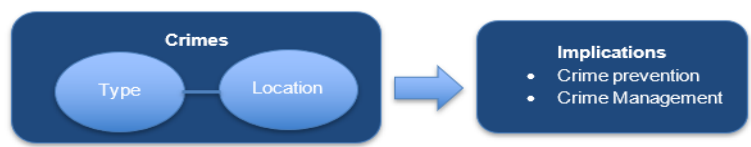

Figure 1. Framework of the Study

- *Research Design and Data collection*

The study used a retrospective quantitative design by utilizing records of crime incidence in the Philippines. The dataset used in this study was obtained from the publication of the Philippine Statistics Authority (PSA), specifically in Chapter 17, which is the Public Order, Safety and Justice Statistics of their Annual Statistical Yearbook. The data was open and free for public use as stipulated in the PSA publication. The dataset contained the volume of index crimes in the Philippines from 2016 to 2020. The index crimes were broken down into two major categories, crimes against persons and crimes against property. Crimes against persons include murder, homicide, rape, and physical injury. On the other hand, crimes against property include theft, robbery, car-napping, and cattle rustling. Incidence of crime by crime type was available for the different administrative regions in the Philippines, namely, National Capital Region (NCR); Cordillera Administrative Region (CAR); Ilocos Region; Cagayan Valley; Central Luzon, Cavite, Laguna, Batangas, Rizal, and Quezon (CALABARZON), among various others.

- *Statistical Analysis*

All categorical variables were presented as numbers (n) and percentages (%). A Chi-square test and correlation plot of chi-square residual was used to determine the associations between the locations and types of index crimes. A correlation plot or dot plot of the chi-square residual was used to investigate the patterns of associations. It is interpreted based on the size and color of dots. Dots are proportional to the magnitude of association. If there is a strong positive association, the dot will appear large dark blue, while a strong negative association will appear to be large dark red. Furthermore, bar charts were used to display the annual rate of change for rape incidence. A two-tailed p-value of < 0.05 was considered statistically significant for all tests. Statistical analysis was carried out using the R Programming language version 4.1.3.

## Results

- *Volume of Index Crimes*

Table 1 shows the incidence of index crimes from 2016 to 2020 broken down into two major categories: crimes against persons and crimes against properties. There has been a steady annual decline of 15% to 28% in the volume of crimes against persons during this period. On the other hand, a steady annual decline of at least 11% and as much as 49% has been observed in the volume of crimes against property. The annual decline rate ranges from 16% to 40% in the overall volume of index crimes.

### Table 1. The volume of index crimes in the Philippines from 2016 to 2020.

| Year | Crimes Against Persons | | | Crimes Against Property | | | Total | | |
|---|---|---|---|---|---|---|---|---|---|
| | Volume | Annual Decline | | Volume | Annual Decline | | Volume | Annual Decline | |
| 2016 | 58925 | | | 80652 | | | 139577 | | |
| 2017 | 50267 | -15% | | 57271 | -29% | | 107538 | -23% | |
| 2018 | 39041 | -22% | | 42372 | -26% | | 81413 | -24% | |
| 2019 | 30690 | -21% | | 37524 | -11% | | 68214 | -16% | |
| 2020 | 22123 | -28% | | 19137 | -49% | | 41260 | -40% | |

The declining volume of index crimes is a piece of evidence that the government's continuing efforts to fight criminality are working. Furthermore, the ability of the police force to solve cases has caused an increase in the crime solution efficiency rating of 78.62 percent. These marked improvements in the overall crime picture translate to a better security outlook among our people and add to upbeat investor confidence that spurs economic growth despite the ongoing health crisis due to the pandemic (Benter & Cawi, 2021; Galabin, Pallega, & Recapente, 2021; Mark & Sarcena, 2021). In addition, it is notable that in 2020, during which the whole country is placed in lockdowns due to COVID-19, there is around a 49% decline in the incidence of crimes against property while only a 28% decline in crime against persons (Interpol, 2020; Payne, Morgan, & Piquero, 2021).

The United Nations Office of Drugs and Crimes (UNODC) noted in their research report in 2020 that crimes against property such as robbery and theft decreased by 50 percent in many countries, particularly those with stricter lockdowns. Furthermore, crimes against the person, such as homicide during lockdown periods, have declined by as much as 25 percent; however, this was short-lived since there is a noticeable increase in homicide rates when lockdowns are lifted (Chainey & Muggah, 2022). The Philippines was placed in a series of lockdowns in 2020, explaining the considerable decline in index crimes.

- *Index Crimes by location*

The distribution of the annual volume of index crimes according to the different administrative regions in the Philippines is presented in Table 2. The bulk of crime incidence is observed in the National Capital Region (16% to 18%), followed by Central Visayas (11% to 15%) and Western Visayas (6% to 10%) from 2016 to 2020. The Bangsamoro Autonomous Region in Muslim Mindanao (BARMM), Cordillera Administrative Region (CAR), Caraga, and MIMAROPA regions reported a low incidence of index crimes.

**26**

Several studies have suggested that population density may explain crime incidence. There are higher opportunities for crime in places where the population density is high (Harries, 2006), particularly in contexts where human crowding influences aggression and hostility (Kvalseth, 1977; Regoeczi, 2003). The National Capital Region is the most densely populated in the Philippines, with a population density of 21,765 persons per square kilometer. The average national population density per square kilometer is only 363 persons. The Cordillera Administrative Region, MIMAROPA, and BARMM have the lowest population densities of 91, 109, and 120 persons per square kilometer.

Table 2. Rate of Decline of Annual Volume of Index Crimes during the pandemic.

| Administratve Regions | Annual Volume of Index Crimes (n%) | | | | | Average Annual Rate of Decline | |
|---|---|---|---|---|---|---|---|
| | 2016 | 2017 | 2018 | 2019 | 2020 | Pre-pandemic | Pandemic |
| NCR | 21681(16%) | 17788(17%) | 14650(18%) | 12313(18%) | 7120(17%) | -17% | -42% |
| CAR | 3990(3%) | 1674(2%) | 1159(1%) | 950(1%) | 579(1%) | -36% | -39% |
| I - Ilocos Region | 4867(3%) | 3712(3%) | 2910(4%) | 2402(4%) | 1481(4%) | -21% | -38% |
| II - Cagayan Valley | 3663(3%) | 2716(3%) | 2389(3%) | 1967(3%) | 1258(3%) | -19% | -36% |
| III - Central Luzon | 10713(8%) | 8640(8%) | 6688(8%) | 5836(9%) | 3589(9%) | -18% | -39% |
| IV-A CALABARZON | 13070(9%) | 11462(11%) | 8312(10%) | 7204(11%) | 4520(11%) | -18% | -37% |
| IV-B MIMAROPA | 2145(2%) | 1718(2%) | 1427(2%) | 1244(2%) | 831(2%) | -17% | -33% |
| V - Bicol Region | 9614(7%) | 7226(7%) | 4934(6%) | 4155(6%) | 2466(6%) | -24% | -41% |
| VI - Western Visayas | 14077(10%) | 11053(10%) | 5030(6%) | 3960(6%) | 2429(6%) | -32% | -39% |
| VII - Central Visayas | 17266(12%) | 12295(11%) | 13637(17%) | 11213(16%) | 6349(15%) | -12% | -43% |
| VIII - Eastern Visayas | 3842(3%) | 3022(3%) | 2894(4%) | 2407(4%) | 1543(4%) | -14% | -36% |
| IX - Zamboanga Peninsula | 6561(5%) | 4936(5%) | 4213(5%) | 3569(5%) | 2114(5%) | -18% | -41% |
| X - Northern Mindanao | 8838(6%) | 6859(6%) | 3266(4%) | 2703(4%) | 1681(4%) | -31% | -38% |
| XI - Davao Region | 6468(5%) | 5020(5%) | 3791(5%) | 3240(5%) | 2052(5%) | -20% | -37% |
| XII - SOCCSKSARGEN | 8117(6%) | 6299(6%) | 3331(4%) | 2764(4%) | 1722(4%) | -29% | -38% |
| XIII - Caraga | 2595(2%) | 1890(2%) | 1856(2%) | 1547(2%) | 999(2%) | -15% | -35% |
| BARMM | 2070(1%) | 1228(1%) | 926(1%) | 740(1%) | 528(1%) | -28% | -29% |

Furthermore, there are notable annual decreases in the annual volume of index crimes in different administrative regions in the Philippines from 2016 to 2020. The average rate of decline ranges from 12% to as much as 39% in all administrative regions. The Cordillera Administrative Region, Western Visayas Region, and Northern Mindanao tallied a decline rate of at least 30%. Central Visayas and Western Visayas have recorded a decline rate of 12% and 14%, respectively. The government's success in reducing crime incidence in the country is attributed to many factors, mainly due to the effective government policies and programs.

The crime incidence has continued to drop in 2020 during the pandemic. Through the recommendation of the Interagency task force against COVID-19 (IATF), which oversees the overall management of COVID-19 in the country, the government has created a categorization that serves as the general guide in the level of mobility restrictions in different regions and localities. The limitations of human mobility during the pandemic have resulted in a significant decline in crime incidence in the different administrative regions in the Philippines. Sixteen (16) administrative regions have at least a 30% rate of annual decline in crime index, while during the pre-pandemic, only one (3) administrative region has achieved the said rate.

27

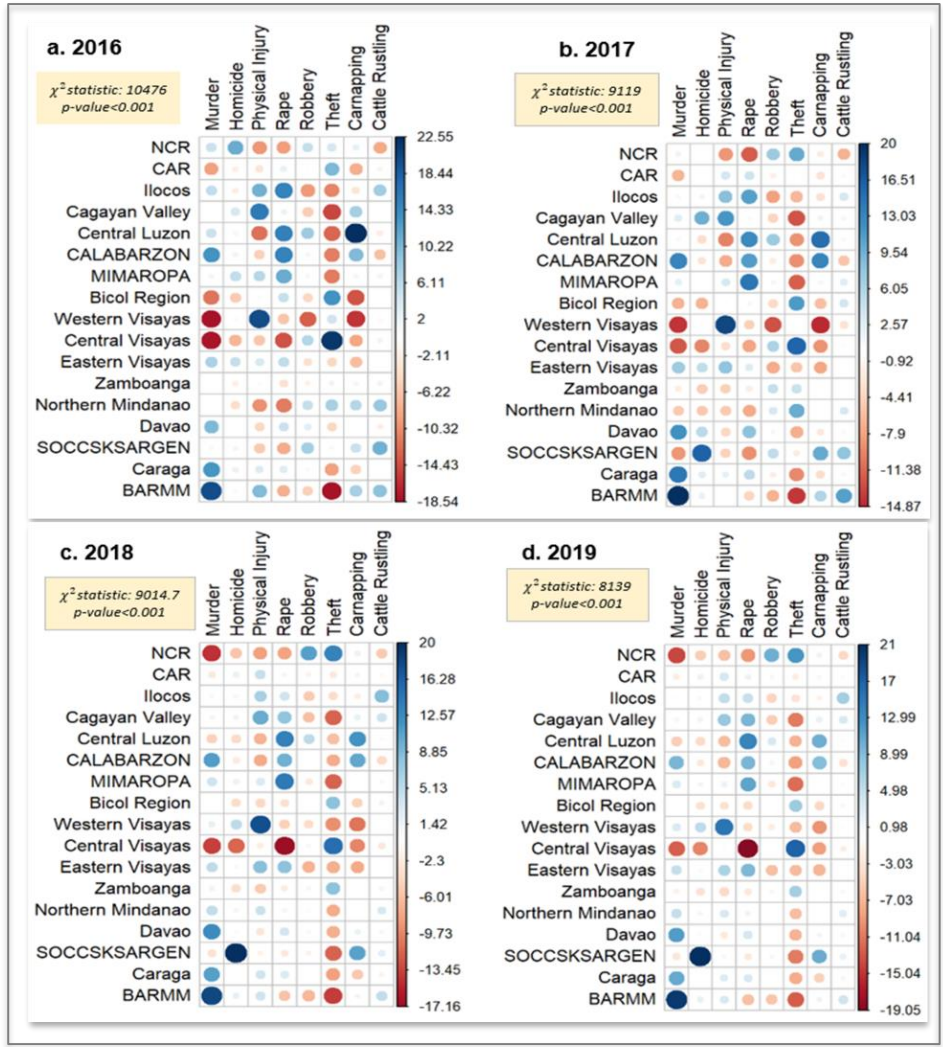

Figure 2. Correlation plot between location and type of index crimes.

- *Associations of Location and Type of Index Crimes*

There is a significant association between the location and type of index crime in the Philippines before the pandemic, from 2016 to 2019 (Figure 2, all p-values<0.001), suggesting that crime incidence varies across locations and certain types of crimes do not randomly occur in different locations. In addition, Figure 2 shows the correlation plot between the type of index crimes and the location in which they were committed before the pandemic. Our results revealed that murder is highly associated with BARMM and moderately associated with Caraga, CALABARZON, and Davao Region. Furthermore, homicide is highly associated with SOCCSKSARGEN, while physical injury is found to have strong associations with Western Visayas and Cagayan Valley. Rape is found to have strong associations with four locations, namely, Ilocos, Central Luzon, CALABARZON, and MIMAROPA. Robbery is associated with NCR, while theft is associated with NCR and Central Visayas. Car-napping is associated with Central Luzon, CALABARZON, and SOCCSKSARGEN, while cattle rustling is associated with Ilocos Region (Table 3).

**28**

A similar analysis was done on the crime incidence in 2020, during which the pandemic hit the whole country and was placed in lockdowns for several periods (Joaquin & Biana, 2021). There is a significant association between location and type of index crimes, which implies that crime incidence of different types during lockdowns varies across different locations. Figure 3 shows the correlation plot between the type and location index crimes during the pandemic. Murder is found to have strong associations with BARMM, Caraga, CALABARZON, and Davao Region, while homicide has a strong association with SOCCSKSARGEN. Physical Injury is strongly associated with Western Visayas, while robbery is associated with NCR and Central Visayas. Furthermore, rape is associated with eleven locations, namely, CAR, Ilocos, Cagayan Valley, Central Luzon, CALABARZON, MIMAROPA, Eastern Visayas, Northern Mindanao, Davao, and CARAGA

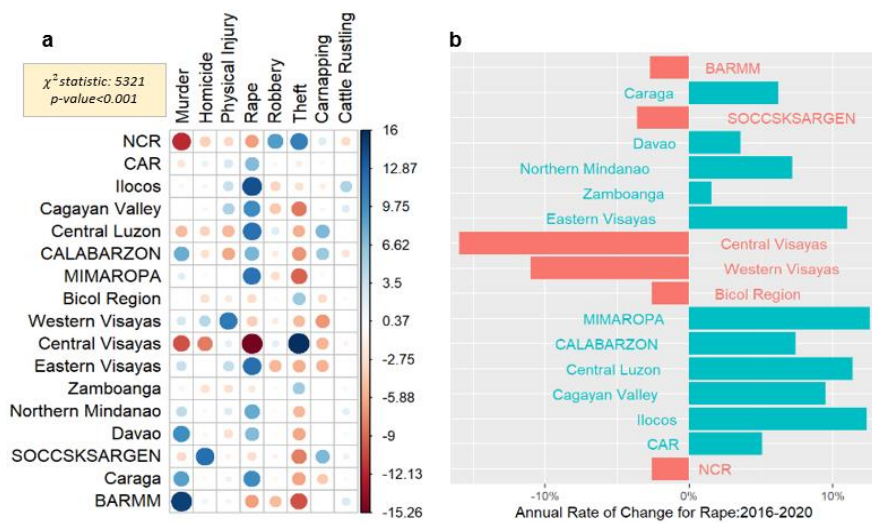

Figure 3. Correlation plot between location and type of index crimes (a), Annual rate of change for Rape from 2016 to 2020

Figure 3 also shows the average rate of change for rape incidence from 2016 to 2020 in different locations in the Philippines. This plot shows that there is indeed an increase in the number of rape incidents in these regions, namely, CAR, Ilocos, Cagayan Valley, Central Luzon, CALABARZON, MIMAROPA, Eastern Visayas, Northern Mindanao, Davao, and CARAGA. On the other hand, rape incidence has declined over the years in the following locations, BARMM, SOCCSKSARGEN, Central Visayas, Western Visayas, Bicol Region, and NCR.

It is notable that the pattern of associations between location and type of index crimes before the pandemic and during the period of the pandemic did not change much for most types of index crimes, namely, murder, homicide, robbery, physical injury, theft, car-napping, and cattle rustling (Table 3). Although a consistent decline is observed, the incidence pattern remains the same for these crime types. On the contrary, a different observation is found in the associations between rape and locations where such crime is committed. Rape is associated with ten (10) locations during the pandemic, while it is only associated with four (4) locations before the pandemic. Rape, therefore, had become a significant crime during the pandemic in

29

many regions of the Philippines. Reports from other countries also stated an increase in sexually related violence during the pandemic. French police reported a nationwide spike of about 30%, while there is an 18% increase in domestic violence and sexual assaults in Spain. The National Alliance to end sexual violence has observed an increase of around 40% of rape cases in most of the rape crisis centers that they have surveyed around the globe.

Table 3. List of locations associated with specific type of index crimes.

| Index Crimes | Pre-pandemic | During pandemic |
| --- | --- | --- |
| Murder | BARMM, Caraga, CALABARZON, Davao Region (4) | BARMM, Caraga, CALABARZON, Davao Region (4) |
| Homicide | SOCCSKSARGEN (1) | SOCCSKSARGEN (1) |
| Physical Injury | Western Visayas and Cagayan Valley (2) | Western Visayas and Cagayan Valley (2) |
| Rape | Ilocos, Central Luzon, CALABARZON, and MIMAROPA (4) | CAR, Ilocos, Cagayan Valley, Central Luzon, CALABARZON, MIMAROPA, Eastern Visayas, Northern Mindanao, Davao, and CARAGA (10) |
| Robbery | NCR (1) | NCR (1) |
| Theft | NCR and Central Visayas (2) | NCR and Central Visayas (2) |
| Carnapping | Central Luzon, CALABARZON, and SOCCSKSARGEN (4) | Central Luzon, CALABARZON, and SOCCSKSARGEN (4) |
| Cattle Rustling | Ilocos (1) | Ilocos (1) |

**Discussion**

Crime prevention and management are at the forefront of the agenda of the Philippine government under the Duterte administration. Its campaign against war on drugs is aimed at reducing criminality and uplifting the lives of the Filipino people. Our study revealed that the continuing efforts of the government to fight criminality have resulted in the increased crime solution efficiency of the police force. The government's success in reducing crime incidence in the country is attributed to many factors, mainly due to the effective government policies and programs. Similar studies have shown that effective government programs significantly reduce crime incidence in different localities (Arvate et al., 2018; Lilley & Boba, 2009; Maguire, Hardy, & Lawrence, 2004).

The crime incidence has continued to drop in 2020 during the pandemic. Through the recommendation of the Interagency task force (IATF) against COVID-19, which oversees the overall management of COVID-19 in the country, the government has created a categorization that serves as the general guide in the level of mobility restrictions in different regions and localities. The limitations of human mobility during the pandemic have resulted in a significant decline in crime incidence in the different administrative regions in the Philippines. Furthermore, the COVID-19 lockdowns in the Philippines have contributed to the decline of crime incidence in the Philippines. Boman and Mowen (2021) noted that global crime trends have declined during the pandemic. According to them, this is expected since during mobility restrictions, opportunities for crime such as robbery, theft and road violence and crimes also decrease.

The significant association between type and location of index crimes in the Philippines suggests that crimes vary across locations and the occurrences of crimes is not random. Rape has become predominant during the pandemic in 11 regions. The findings are similar to many studies wherein crimes against women such as rape, domestic violence and sexual abuse have become prominent during COVID-19 (Rapee et al., 2022; Rockowitz et al., 2021; Sifat, 2020). All these studies consistently showed that rape had become a significant crime during the pandemic in many regions of the Philippines. Reports from other countries also stated an increase in sexually related violence during the pandemic. French police reported a nationwide spike of about 30%, while there was an 18% increase in domestic violence and sexual assaults in Spain. The National Alliance to end sexual violence observed an increase of around 40% of rape cases in most of the rape crisis centers that were surveyed around the globe. Furthermore, the pattern of associations between location and type of index crimes before and after the pandemic did not change much for most crime types, namely, murder, homicide, robbery, physical injury, theft, car-napping, and cattle rustling (Muldoon et al., 2021; Walker, 2020).

**Conclusion**

The continuing effort of the government to fight against criminality has resulted in a steady decline in the incidence of index crimes in the Philippines. The pandemic has also contributed to the decline of crime incidence in the country. In terms of location, the incidence of index crimes is high in densely populated areas. Hence, it is recommended to increase police presence and surveillance activities in highly populated areas. Furthermore, our analysis revealed that crime type is indeed associated with the location. Crime types that should be prioritized by location are identified in this study. Hence, regional government executives can use our results as input to policy and programs aimed at crime prevention and management in their localities. Moreover, rape had become a severe issue in many locations during the pandemic. Thus, there must be a specific intervention to address sexual violence during community lockdowns.

**Acknowledgement**

This study was developed with ESF Project No. 8.2.2.0/20/I/003 "Strengthening of Professional Competence of Daugavpils University Academic Personnel of Strategic Specialization Branches 3rd Call". The authors thank Analyn Cabras, director of the Coleoptera Research Center of the University of Mindanao for the support and cooperation.

**References**
Arvate, P., Falsete, F. O., Ribeiro, F. G., & Souza, A. P. (2018). Lighting and homicides: Evaluating the effect of an electrification policy in rural Brazil on violent crime reduction. *Journal of quantitative criminology, 34*(4), 1047-1078. https://doi.org/10.1007/s10940-017-9365-6
Asher, J., & Horwitz, B. (2020). How do the Police actually spend their time? *The New York Times, 19*. https://www.courts.ca.gov/opinions/links/S249792-LINK7.PDF
Balmori de la Miyar, J. R., Hoehn-Velasco, L., & Silverio-Murillo, A. (2021). The U-shaped crime recovery during COVID-19: evidence from national crime rates in Mexico. *Crime science, 10*(1), 1-23. https://doi.org/10.1186/s40163-021-00147-8

Benter, J. B., & Cawi, R. D. (2021). The State of the Art of the Philippine National Police Crime Laboratory Services. *Available at SSRN 3826806*. https://dx.doi.org/10.2139/ssrn.3826806

Blau, J. R., & Blau, P. M. (1982). The cost of inequality: Metropolitan structure and violent crime. *American sociological review, 47*(1), 114-129. https://doi.org/10.2307/2095046

Boman, J. H., & Gallupe, O. (2020). Has COVID-19 changed crime? Crime rates in the United States during the pandemic. *American journal of criminal justice, 45*(4), 537-545. https://doi.org/10.1007/s12103-020-09551-3

Boman, J. H., & Mowen, T. J. (2021). Global crime trends during COVID-19. *Nature Human Behaviour, 5*(7), 821-822. https://doi.org/10.1038/s41562-021-01151-3

Buil-Gil, D., Zeng, Y., & Kemp, S. (2021). Offline crime bounces back to pre-COVID levels, cyber stays high: interrupted time-series analysis in Northern Ireland. *Crime science, 10*(1), 1-16. https://doi.org/10.1186/s40163-021-00162-9

Chainey, S., & Muggah, R. (2022). Homicide concentration and retaliatory homicide near repeats: An examination in a Latin American urban setting. *The Police Journal, 95*(2), 255-275. https://doi.org/10.1177/0032258x20980503

Chiricos, T. G. (1987). Rates of crime and unemployment: An analysis of aggregate research evidence. *Social problems, 34*(2), 187-212. https://doi.org/10.2307/800715

Department of Budget and Management. (2018). President Duterte fulfills campaign promise, doubles salaries of cops, soldiers. https://dbm.gov.ph/index.php/secretary-s-corner/press-releases/list-of-press-releases/425-president-duterte-fulfills-campaign-promise-doubles-salaries-of-cops-soldiers

Fowles, R., & Merva, M. (1996). Wage inequality and criminal activity: An extreme bounds analysis for the United States, 1975–1990. *Criminology, 34*(2), 163-182. https://doi.org/10.1111/j.1745-9125.1996.tb01201.x

Galabin, N. D., Pallega, R. B., & Recapente, M. A. (2021). Philippine National Police Initiatives in the Promotion of Peace and Order in Iligan City: Basis for Policy Recommendation. *International Journal of Multidisciplinary: Applied Business and Education Research, 2*(9), 773-785. https://doi.org/10.11594/ijmaber.02.09.08

Gita-Carlos, R. A. (2019). *Duterte's relentless war on drugs, corruption, crime*. Philippine News Agency. https://www.pna.gov.ph/articles/1089333

Harries, K. (2006). Property crimes and violence in United States: an analysis of the influence of population density. *International Journal of Criminal Justice Sciences, 1*(2), 24–34. http://www.sascv.org/ijcjs/harries.pdf

Hilsenrath, P. E. (2021). Ethics and Economic Growth in the Age of COVID-19: What Is a Just Society to Do? *The Journal of Rural Health, 37*(1), 146-147. https://doi.org/10.1111/jrh.12434

Interpol. (2020). *News and Events*. https://www.interpol.int/en/News-and-Events

Irvin-Erickson, Y., & La Vigne, N. (2015). A spatio-temporal analysis of crime at Washington, DC metro rail: Stations' crime-generating and crime-attracting characteristics as transportation nodes and places. *Crime science, 4*(1), 1-13. https://doi.org/10.1186/s40163-015-0026-5

Joaquin, J. J. B., & Biana, H. T. (2021). Philippine crimes of dissent: Free speech in the time of COVID-19. *Crime, Media, Culture, 17*(1), 37-41. https://doi.org/10.1177/1741659020946181

Kvalseth, J. O. (1977). A note on the effects of population density and unemployment on urban crime. *Criminology, 15*(1), 105-110. https://doi.org/10.1111/j.1745-9125.1977.tb00051.x

Leong, K., & Sung, A. (2015). A review of spatio-temporal pattern analysis approaches on crime analysis. *International E-Journal of Criminal Sciences, 9*, 1-33. https://glyndwr.repository.guildhe.ac.uk/id/eprint/8255

Lilley, D., & Boba, R. (2009). Crime reduction outcomes associated with the State Criminal Alien Assistance Program. *Journal of Criminal Justice, 37*(3), 217-224. https://doi.org/10.1016/j.jcrimjus.2009.04.001

Lusthaus, C., Adrien, M., Anderson, G., & Carden, F. (1999). *Enhancing Organizational Performance: A Toolbox for Self-Assessment*. International Development Research Centre. Ottawa, ON, Canada. http://lib.icimod.org/record/10368/files/1377.pdf

Maguire, S., Hardy, C., & Lawrence, T. B. (2004). Institutional entrepreneurship in emerging fields: HIV/AIDS treatment advocacy in Canada. *Academy of management journal, 47*(5), 657-679. https://doi.org/10.5465/20159610

Mark, P., & Sarcena, J. D. G. (2021). Police Operational Activities and Crime Commission in a City in the Philippines. *IOER International Multidisciplinary Research Journal, 2*(1), 79-88. https://www.ioer-imrj.com/wp-content/uploads/2021/04/Police-Operational-Activities-and-Crime-Commission-in-a-City-in-the-Philippines.pdf

McCarthy, L. P. (2020). A stranger in strange lands: A college student writing across the curriculum. In *Landmark Essays* (pp. 125-155). Routledge. https://www.taylorfrancis.com/chapters/edit/10.4324/9781003059219-12

Meyer, B. H., Prescott, B., & Sheng, X. S. (2022). The impact of the COVID-19 pandemic on business expectations. *International Journal of Forecasting, 38*(2), 529-544. https://doi.org/10.1016/j.ijforecast.2021.02.009

Muldoon, K. A., Denize, K. M., Talarico, R., Fell, D. B., Sobiesiak, A., Heimerl, M., & Sampsel, K. (2021). COVID-19 pandemic and violence: rising risks and decreasing urgent care-seeking for sexual assault and domestic violence survivors. *BMC medicine, 19*(1), 1-9. https://doi.org/10.1186/s12916-020-01897-z

Newton, A., & Felson, M. (2015). Crime patterns in time and space: The dynamics of crime opportunities in urban areas. *Crime science, 4*(1), 1-5. https://doi.org/10.1186/s40163-015-0025-6

Nivette, A. E., Zahnow, R., Aguilar, R., Ahven, A., Amram, S., Ariel, B., Burbano, M. J. A., Astolfi, R., Baier, D., & Bark, H.-M. (2021). A global analysis of the impact of COVID-19 stay-at-home restrictions on crime. *Nature Human Behaviour, 5*(7), 868-877. https://doi.org/10.1038/s41562-021-01139-z

Numbeo. (n.d.). *Crime Index by Country 2022 Mid-Year*. https://www.numbeo.com/crime/rankings_by_country.jsp

Payne, J. L., Morgan, A., & Piquero, A. R. (2021). Exploring regional variability in the short-term impact of COVID-19 on property crime in Queensland, Australia. *Crime science, 10*(1), 1-20. https://doi.org/10.1186/s40163-020-00136-3

Rapee, R., Wignall, A., Spence, S., Cobham, V., & Lyneham, H. (2022). *Helping your anxious child: A step-by-step guide for parents*. New Harbinger Publications. https://www.newharbinger.com/9781684039913/helping-your-anxious-child/

Regoeczi, W. C. (2003). When context matters: A multilevel analysis of household and neighbourhood crowding on aggression and withdrawal. *Journal of environmental Psychology, 23*(4), 457-470. https://doi.org/10.1016/S0272-4944(02)00106-8

Rockowitz, S., Stevens, L. M., Rockey, J. C., Smith, L. L., Ritchie, J., Colloff, M. F., Kanja, W., Cotton, J., Njoroge, D., & Kamau, C. (2021). Patterns of sexual violence against adults and children during the COVID-19 pandemic in Kenya: a prospective cross-sectional study. *BMJ open, 11*(9), e048636. http://dx.doi.org/10.1136/bmjopen-2021-048636

Sifat, R. I. (2020). Sexual violence against women in Bangladesh during the COVID-19 pandemic. *Asian journal of psychiatry, 54*, 102455. https://doi.org/10.1016/j.ajp.2020.102455

Statista. (2022). *Number of crimes in the Philippines against property by type of index crimes*. https://www.statista.com/statistics/1170818/philippines-number-crimes-against-property-by-type-of-index-crime/

Struett, D. (2020). *Chicago gun violence still up 50% through end of October as other crime falls*. Chicago Sun Times. https://chicago.suntimes.com/crime/2020/11/1/21544510/chicago-gun-violence-statistics-homicide-shooting-cpd-police

Timberman, D. (2019). *Philippine Politics Under Duterte: A Midterm Assessment*. Carnegie Endowment for international peace. https://carnegieendowment.org/2019/01/10/philippine-politics-under-duterte-midterm-assessment-pub-78091

Walker, T. (2020). *A Second, Silent Pandemic: Sexual Violence in the time of COVID-19*. Harvard medical school primary care review. https://info.primarycare.hms.harvard.edu/review/sexual-violence-and-covid

Zhou, B., Chen, L., Zhao, S., Zhou, F., Li, S., & Pan, G. (2021). Spatio-temporal analysis of urban crime leveraging multisource crowdsensed data. *Personal and Ubiquitous Computing*, 1-14. https://doi.org/10.1007/s00779-020-01456-6